\documentclass[12pt]{iopart}
\usepackage{epsfig}

\graphicspath{{/.}{/home/scratch/konrad/project/kaidalov/GluonShadowing/pictures/}}
\DeclareGraphicsExtensions{.eps}

\newcommand{\beq}{\begin{eqnarray}}
\newcommand{\eeq}{\end{eqnarray}}
\newcommand{\be}{\begin{eqnarray*}}
\newcommand{\ee}{\end{eqnarray*}}

\def\lsim{\raise0.3ex\hbox{$<$\kern-0.75em\raise-1.1ex\hbox{$\sim$}}}
\def\gsim{\raise0.3ex\hbox{$>$\kern-0.75em\raise-1.1ex\hbox{$\sim$}}}

\begin{document}

\title[]
{Multiplicity and cold-nuclear matter effects from Glauber-Gribov
  theory at LHC}

\author{I.C. Arsene$^{1}$, L. Bravina$^{1,2}$, A.B. Kaidalov$^{3}$, K.
  Tywoniuk$^{1}$ and E. Zabrodin$^{1,2}$}

\address{$^1$ Department of Physics, University of Oslo, N-0316 Oslo}
\address{$^2$ Institute of Nuclear Physics, Moscow State University,
  RU-119899 Moscow}
\address{$^3$ Institute of Theoretical and Experimental Physics,
  RU-117259 Moscow}
\ead{konrad@fys.uio.no}
 
\begin{abstract}
We present predictions for nuclear modification factor in proton-lead
collisions at LHC energy 5.5 TeV from Glauber-Gribov theory of
nuclear shadowing. We
have also made predictions for baseline cold-matter nuclear effects in
lead-lead collisions at the same energy.
\end{abstract}

\maketitle

\section{Introduction}
The system formed in nucleus-nucles (AA)
collisions at LHC will provide further insight into the
dynamics of the deconfined state of nuclear matter. There are also 
interesting effects anticipated for the initial state of the incoming
nuclei related 
to shadowing of nuclear parton distributions and the space-time
picture of the interaction. These should be studied in the more
``clean'' environment of a proton-nucleus collision. The initial-state effects
constitute a baseline for calculation of the density of particles at
all rapidities and affect therefore also high-$p_\bot$ particle
suppression and jet quenching, as well as the total multiplicity.

Both soft and relatively high-$p_\bot$, $p_\bot < 10$ GeV/c, particle
production in pA at LHC energies probe the low-$x$ 
gluon distribution of the target nucleus at moderate scales, $Q^2
\sim p_\bot^2$, and is therefore mainly influenced by nuclear shadowing. In
the Glauber-Gribov theory \cite{Gri69}, shadowing at low-$x$ is
related to diffractive structure functions of the nucleon, which are
studied at HERA. The space-time picture 
of the interaction is altered from a longitudinally ordered
rescattering at low energies, to a coherent interaction of the
constituents of the incoming wave-functions at high energy.
Shadowing affects both soft and hard processes. Calculation of gluon
shadowing was performed in our recent paper \cite{Ars07}, where Gribov
approach for the calculation of nuclear structure functions was used.
The Schwimmer model was used to account for higher-order
rescatterings. The gluon diffractive distributions are 
taken from the most recent experimental parameterizations \cite{H106}.

\section{Particle production at LHC}
\begin{figure}[t!]
  \begin{center}
    \includegraphics[scale=.5]{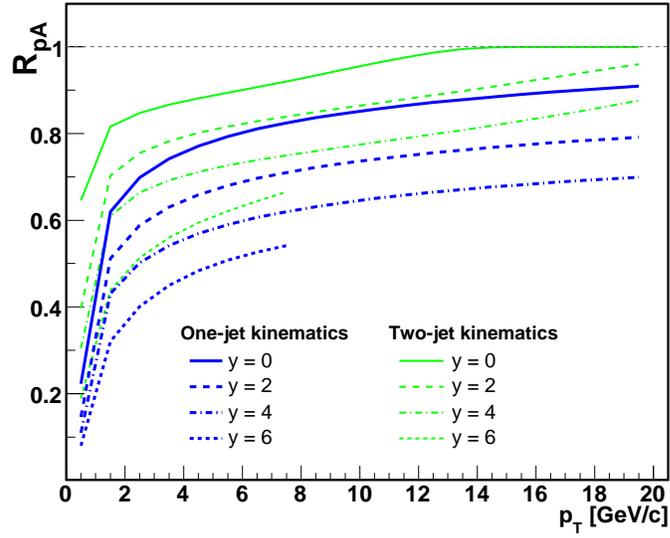}
  \end{center}
  \caption{Shadowing as a function of transverse momentum for p+Pb collisions
    at $\sqrt{s}$ = 5.5 TeV.}
  \label{fig:R_pA}
\end{figure}

Shadowing will lead to a suppression both at mid- and forward rapidities in
p+Pb collisions at $\sqrt{s} = 5.5$ TeV as seen in
Fig.~\ref{fig:R_pA}. We have plotted the curves for two
distinct kinematical scenarios of 
particle production; one-jet kinematics which may be well motivated for
particle production at $p_\bot < 2$ GeV/c and two-jet kinematics that
apply for high-$p_\bot$ particle producion. The uncertainty in the
curves is due to uncertainty in the parameterization of gluon
diffractive distribution functions. Cronin effect is not included in
the curves of Fig.~\ref{fig:R_pA}. We estimate it to be a 10\% effect
at these energies.

\begin{figure}[t!]
  \begin{minipage}[t]{1.\linewidth}
    \begin{center}
      \includegraphics[scale=.5]{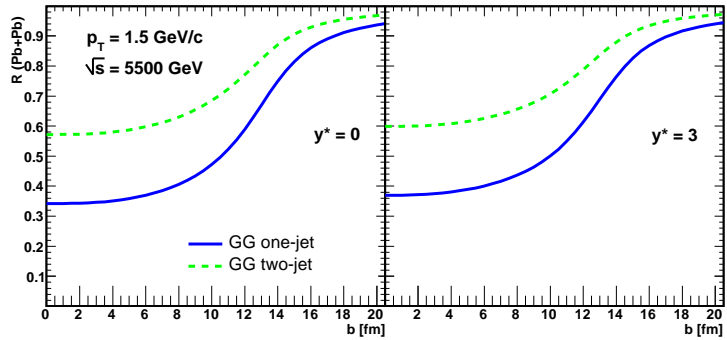}\\
      \includegraphics[scale=.4]{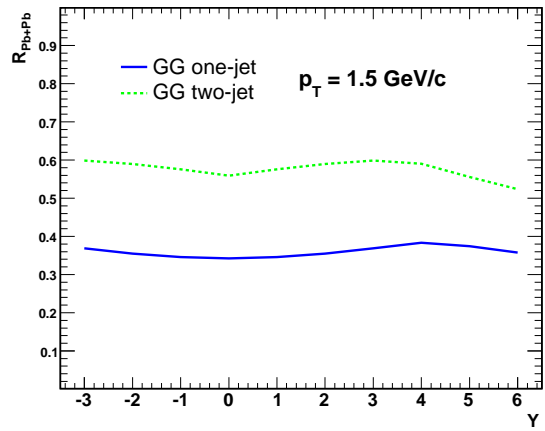}
    \end{center}
  \end{minipage}
  \caption{Shadowing as a function of centrality (top) and rapidity
    (bottom) for Pb+Pb collisions
    at $\sqrt{s}$ = 5.5 TeV.}
  \label{fig:ColdNucl}
\end{figure}

In Fig.~\ref{fig:ColdNucl} we present the suppresion due to
cold-nuclear effects in Pb+Pb collisions at $\sqrt{s} = 5.5$ TeV as a
function of centrality (top) and rapidity (bottom). Also here we
present the results for two kinematics.


\end{document}